\documentclass[nonatbib,preprint,12pt]{elsarticle}

\usepackage{amssymb}
\usepackage{amsmath}

\makeatletter
\let\c@author\relax
\makeatother

\usepackage{xcolor}
\usepackage{textgreek}
\usepackage{booktabs}
\usepackage{tikz, tikz-qtree}
\usepackage{listings}
\usepackage{xcolor}
\usepackage{float}
\usepackage{multicol}
\usepackage[sorting=none]{biblatex}
\usepackage{comment}
\usepackage{soul}

\journal{NIM A}

\begin{document}

\newcommand{\redbold}[1]{
    \textcolor{red}{\textbf{#1}}
}
\newcommand{\bluebold}[1]{
    \textcolor{blue}{\textbf{#1}}
}
\newcommand{\greenbold}[1]{
    \textcolor{teal}{\textbf{#1}}
}
\newcommand{\orangebold}[1]{
    \textcolor{orange}{\textbf{#1}}
}

\definecolor{bkgnd}{rgb}{0.97, 0.99, 1}
\definecolor{string}{rgb}{0.00, 0.3, 0.2}

\lstdefinestyle{script}{
    backgroundcolor=\color{bkgnd},   
    commentstyle=\color{blue},
    keywordstyle=\color{teal},
    numberstyle=\tiny\color{gray},
    stringstyle=\color{string},
    basicstyle=\scriptsize\ttfamily,
    breakatwhitespace=false,         
    breaklines=true,                 
    captionpos=b,                    
    keepspaces=true,                                 
    showspaces=false,                
    showstringspaces=false,
    showtabs=false,                  
    tabsize=2
}

\lstset{style=script}

\begin{frontmatter}



\title{SNAPPY CubeSat Control Script Generation and Data File Processing}


\author[inst1]{E. Bierens\fnref{eBier}\corref{cor1}}
\fntext[eBier]{
    \textit{Email:} eabierens@shockers.wichita.edu 
    \textit{Mailing address:} Edward Bierens, Wichita State University, Campus box 32, 1845 Fairmount St., Wichita, KS 67260 
    \textit{Telephone:} (316) 978-3991
}
\author[inst1]{J. Folkerts}
\author[inst1]{B. Doty}
\author[inst1]{H. Meyer\fnref{hMeyer}}
\author[inst1]{N. Solomey\fnref{nSolo}}
\cortext[cor1]{Corresponding Author}

\affiliation[inst1]{organization={Wichita State University},
            addressline={1845 Fairmount St.}, 
            city={Wichita},
            postcode={67260}, 
            state={Kansas},
            country={United States of America}}

\begin{abstract}
This is a document discussing the creation and usage of a server system dedicated to retrieving, processing, and storing data generated from the Solar Neutrino and Astro-Particle PhYsics (SNAPPY) CubeSat by the \textnu{}SOL Project. On a traditional desktop computer with CERN's ROOT and PostgreSQL software installed, and with a file system on two mirrored drives, it is possible to automatically process and organize incoming data, along with keeping a database to record each incoming file along with a command record. In addition to this, an application was created to provide a Graphical User Interface to assist with creating commands to communicate with the CubeSat. With that said, there are still plenty of plans to improve the software, mainly providing an automatic emailing system to notify team members when they are not around the server.
\end{abstract}

\begin{highlights}
\item Set up an automated server system to process and decode detector data and telemetry.

\item A script generator was made to assist with script generation on a satellite.
\end{highlights}


\end{frontmatter}




\section{Introduction}\label{sec:Intro}
The two pieces of software that this paper will discuss is in relation to the \textnu{}SOL Collaboration's mission to send a satellite into near-solar orbit to collect neutrino data closer to the Sun to get improved data as compared to building larger detectors on Earth \cite{ref:nusolmission}. A CubeSat mission will be conducted with the Solar Neutrino and Astro-Particle PhYsics (SNAPPY) CubeSat to measure the in situ rate of background radiation in Low Earth Orbit over the Van-Allen Radiation Belts \cite{ref:nusolmission}.

For this mission, we have prepared a computer server at Wichita State University's campus to automatically sort and decode incoming files; recording them into a database for keeping track of large datasets. The raw data files are parsed into the format used by CERN's ROOT software \cite{ref:cernroot} among other files that are formatted for use in the mission and analysis.

In addition, for operations conducted during the mission, a piece of software was developed to automatically generate CubeSat script files and commands with the proper formatting.


\section{File Types}\label{sec:FileTypes}
The daemon installed on the server looks for, parses, and organizes several types of files. Some of the most important categories include the raw binary data collected from the detector, script logs from CubeSat operations, telemetry data from the satellite, and timestamp files for different mission operation commands. A complete list of file categories along with their associated file extensions can be found within Table \ref{table:types}.

\begin{table}[H]
\centering
\begin{center}
\resizebox{\textwidth}{!}{%
\begin{tabular}{@{}cccc@{}}
\toprule
\textit{\textbf{File Extension}} & \textit{\textbf{Description}}                            & \textit{\textbf{Variable Name}} & \textit{\textbf{File Destination}} \\ \midrule
\textit{.dat}                       & Detector data taken from satellite    & \textit{DETECTOR\_SATELLITE}  & /data/SnappyRuns/RAW \\
\textit{.log}                    & Detector data taken from Marshall's software & \textit{DETECTOR\_GROUND}  &  /data/SnappyRuns/RAW   \\
\textit{.log}                       & Logs from CubeSat interactions & \textit{NANO\_LOG}    &  /data/SnappyRuns/LOGSAT         \\
\textit{.nmcs}                      & Scripts uploaded to the CubeSat                & \textit{SCRIPT\_EXECUTABLE}  &  /data/SnappyRuns/SE/SCRIPTS   \\
\textit{.txt}                       & Logs from the CubeSat scripts  & \textit{SCRIPT\_LOG}   &  /data/SnappyRuns/SE/LOGS        \\
\textit{.bin}                       & CubeSat subsystem telemetry files      & \textit{NANO\_TEL\_DATA}   &  /data/SnappyRuns/TM/BIN    \\
\textit{.mcf}                       & Scripts for mission control      & \textit{MISSION\_CONTROL}   &  /data/SnappyRuns/MCF    \\
\textit{.time}                       & Collection of UNIX timestamps      & \textit{TIME\_COLLECTION}   &  /data/SnappyRuns/TIME    \\
\textit{.lock}                       & Lock file indicating file is being processed       & \textit{LOCK\_FILE}   &  Not Applicable    \\
\textit{{[}unknown/no extension{]}} & Other or no file extension provided                    & \textit{FILE\_UNKNOWN}  &  /data/SnappyRuns/WHAT       \\ \bottomrule
\end{tabular}%
}
\end{center}
\caption{List of all possible file types, the file extension associated with each type, variable name given in the program, and the destination.}
\label{table:types}
\end{table}


\section{Structure of the Database}\label{sec:SysLayout}
To set up the server, a Dell Precision 7920 running AlmaLinux 9.5 was used. The database is powered by PostgreSQL 13.20 \cite{ref:postgresql}, and the detector data is parsed into ROOT 6.32.10. This software utilizes a Redundant Array of Independent Disks Type 1 (RAID 1) to ensure that the database and software are backed up. A script was also created for rapid deployment on a new system in the event of hardware failure during the mission.

One of the database's functions is to record critical information about each file. In general, this includes a file identifier, file version, upload time, comments, and a flag for whether the file has been backed up. Detector data files include scientifically relevant information such as the number of events in a file and the pole from which the data was taken. Log files, telemetry files, and scripts provide the relevant subsystem that they came from, however telemetry files will also include the number of entries the data contains. 

Another function of the database is to keep a log as to what operations were run throughout the mission. One table will contain a list of these operations, where each operation is given an action, the timestamp the action starts at, and the location the action takes place. Two lookup tables are also provided for an easy translation between numeric identifiers and descriptions; one for all possible actions, and one for all possible locations for said actions.

\begin{figure}[H]
\begin{center}
\resizebox{\textwidth}{!}{
    \begin{tikzpicture}[grow=right]
    \tikzset{level distance=150pt, sibling distance=5pt}
    \Tree
    [ 
        .\redbold{/data} 
        \redbold{FittedHists} 
        \redbold{NanoStructs} 
        \redbold{Launchpad} 
        [ 
            .\bluebold{SnappyRuns} 
            \bluebold{LOGSAT} 
            \bluebold{CSV} 
            \bluebold{MCF}
            \bluebold{PNG} 
            \bluebold{RAW} 
            \bluebold{ROOT} 
            [ 
                .\bluebold{SE} 
                \bluebold{LOGS} 
                \bluebold{SCRIPTS} 
            ] 
            [ 
                .\bluebold{TM} 
                \bluebold{UNKNOWN} 
                \bluebold{BIN} 
                [ 
                    .\bluebold{COMM} 
                    \bluebold{BEACON} 
                    \bluebold{GENERAL} 
                    \bluebold{PROTOCOL} 
                    \bluebold{STARTUP} 
                ]
                [ 
                    .\bluebold{EPS} 
                    \bluebold{BEACON} 
                    \bluebold{GENERAL} 
                    \bluebold{PROTOCOL} 
                    \bluebold{STARTUP} 
                ] 
                [ 
                    .\bluebold{FC} 
                    \bluebold{BEACON} 
                    \bluebold{GENERAL} 
                    \bluebold{PROTOCOL} 
                    \bluebold{STARTUP} 
                    \bluebold{ADCS}
                ] 
                [ 
                    .\bluebold{PC} 
                    \bluebold{BEACON} 
                    \bluebold{GENERAL} 
                    \bluebold{PROTOCOL} 
                    \bluebold{STARTUP} 
                ]  
            ] 
            \bluebold{TIME}
            \bluebold{TXT} 
            \bluebold{WHAT}
        ]
        \textcolor{red}{log} 
        \textcolor{red}{sentry}  
    ]

    \end{tikzpicture}
}  
\caption{Structure of the file-system used to store different files from the mission. Nodes with bold text indicate folders, otherwise they indicate a file. Red indicates parts that can only be accessed via elevated privileges, whereas blue indicates parts that can be freely accessed but not modified.}\label{fig:filesystem}
\end{center}
\end{figure}
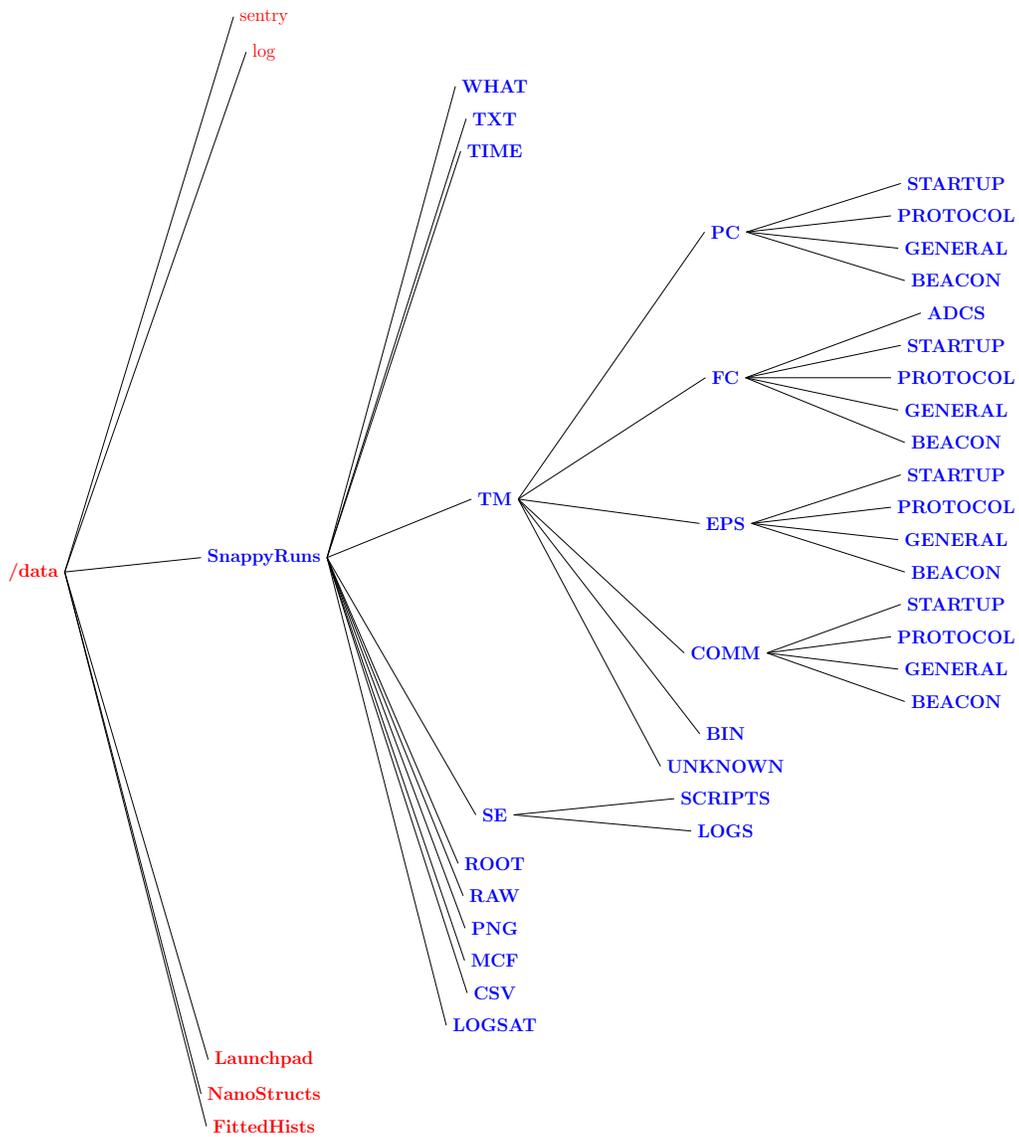

The files within the database reside in a dedicated directory within the RAID 1 drives. The base directory \texttt{/data} contains various files and directories for the daemon's operation, and attempts to modify the contents within this structure requires elevated permissions to preserve data integrity. With that said, general users are allowed to read what is located within the folder \texttt{/data/SnappyRuns} for easy access to the raw and parsed detector data in addition to the various telemetry data from the CubeSat. For full reference, the file system layout can found in Figure \ref{fig:filesystem}.

\begin{figure}[H]
\begin{center}
\resizebox{\textwidth}{!}{
    \includegraphics[]{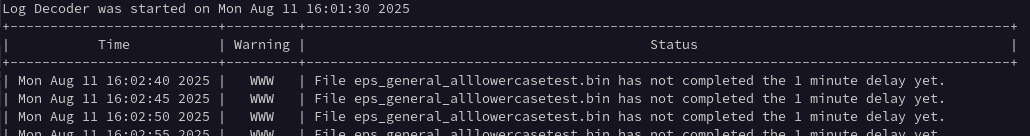}
}
\caption{Screenshot of the server log when the daemon starts}\label{fig:logsample}
\end{center}
\end{figure}

Finally, to keep track of the daemon's actions, a log file exists in \texttt{/data/log}. This file includes the status, warning messages, and error messages in addition to what time said messages have occurred. A sample of the log file has been provided in Figure \ref{fig:logsample}.

\section{Decoding Process \& ROOT Output}\label{sec:Process}
\begin{figure}[H]
\begin{center}
\resizebox{\textwidth}{!}{
    \includegraphics[]{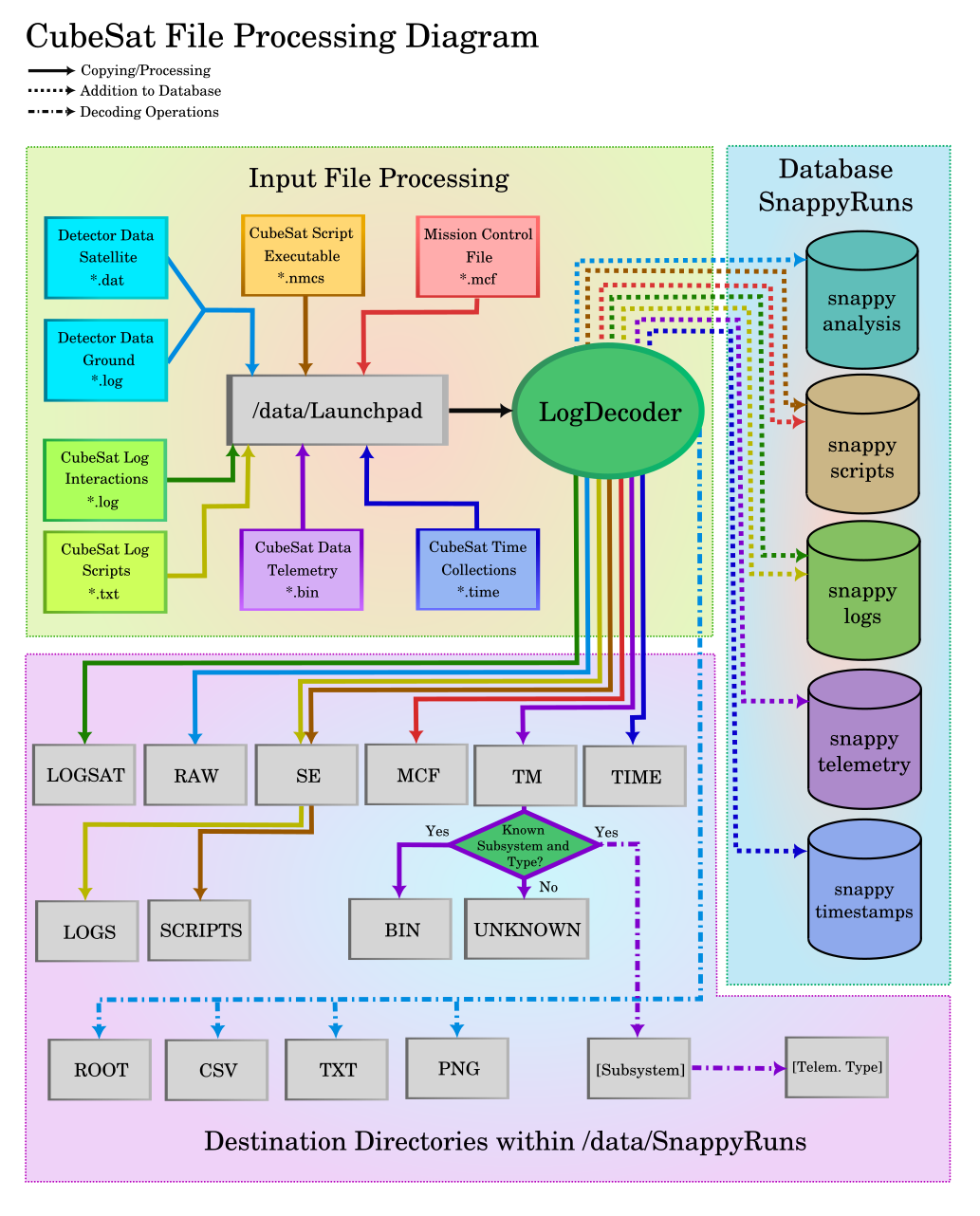}
}
\end{center}
\caption{Visual diagram to better explain how the daemon processes files in \texttt{/data/Launchpad} and how files are output in \texttt{/data/SnappyRuns}.}\label{fig:processdiagram}
\end{figure}

The design of the server is to require as little human intervention as possible. Any of the data, log, or telemetry files are sent to \texttt{/data/Launchpad} and are given a delay of one minute to ensure that there was enough time to copy the file over. Once this delay passes, it is assumed these files are finished copying and are decoded and/or processed by the daemon; they are moved to their appropriate folders, and a record of said file is added into the database. For more information, Figure \ref{fig:processdiagram} provides a complete visualization as to how the types of files in Table \ref{table:types} are processed and stored.

\begin{figure}[H]
\begin{center}
\resizebox{0.5\textwidth}{!}{
    \includegraphics[]{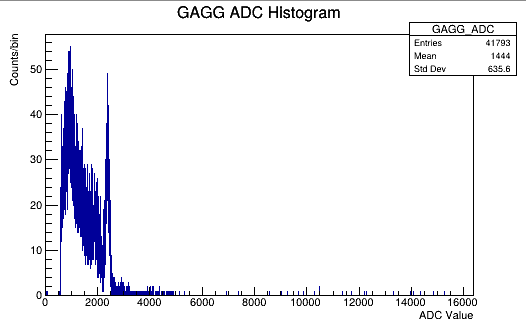}
}
\end{center}
\caption{Generated GAGG histogram contents}\label{fig:rootgagg}
\end{figure}

\begin{figure}[H]
\begin{center}
\resizebox{0.4\textwidth}{!}{
    \includegraphics[]{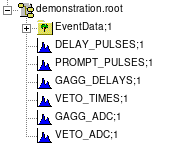}
}
\end{center}
\caption{ROOT file structure of any detector data}\label{fig:rootstruct}
\end{figure}

Previously mentioned in Section \ref{sec:Intro}, one of the most important outputs of the daemon are parsed detector data files designed to be opened with ROOT. Contained in these ROOT files are a TTree data structure along with six histograms. It is important to specify that there are actually two detectors; one utilizing a Gadolinium-Aluminum Gallium Garnet (GAGG) crystal to detect neutrinos, and one for charged particles called the Veto. As such, the TTree data structure called \texttt{EventData} contains the raw information taken from the detector, and a full list of the various data types can be found in Table \ref{table:ROOT}. The six histograms that are provided help automate some of the processes personnel go through to generate the data, where the full functions of which are described in Table \ref{table:Hists}. Finally, Figure \ref{fig:rootgagg} shows a preview as to what the generated histogram contained in \texttt{GAGG\_ADC} looks like (a Zinc-65 source was used as an example), and a preview as to what each ROOT file looks like in ROOT is found in Figure \ref{fig:rootstruct}.

\begin{table}[H]
\centering
\resizebox{\textwidth}{!}{%
\begin{tabular}{@{}ccc@{}}
\toprule
\textit{\textbf{Branch}}   & \textit{\textbf{Data Type}} & \textit{\textbf{Meaning}}                                  \\ \midrule
\textit{TelemetryUNIXTime} & UInt\_t                     & \textit{Raw 32-bit UNIX Epoch Time used in Timestamp.}     \\
\textit{EventTick}         & UShort\_t                   & \textit{16-bit time value of unknown quantity.}            \\
\textit{Mode}              & UChar\_t                    & \textit{CubeSat mode that the detector data was taken in.} \\ 
\textit{GAGG\_RBA\_n\_ADC}    & UShort\_t                   & \textit{Collection of ADC values within GAGG at the index n; ranges from 1 to 5.} \\
\textit{GAGG\_RBA\_n\_Time}   & UShort\_t                   & \textit{Corresponding time values for GAGG ADC values at n.}      \\
\textit{VETO\_RBA\_n\_ADC}    & UShort\_t                   & \textit{Collection of ADC values within the Veto at the index n.} \\
\textit{VETO\_RBA\_n\_Time}   & UShort\_t                   & \textit{Corresponding time values for Veto ADC values at n.}      \\ 
\textit{VETO\_Triggers}       & UShort\_t                   & \textit{Number of triggers in the Veto.}             \\
\textit{VETO\_ExpiredSamples} & UShort\_t                   & \textit{Number of expired samples in the Veto.}      \\
\textit{GAGG\_Triggers}       & UShort\_t                   & \textit{Number of triggers in the GAGG.}             \\
\textit{GAGG\_ExpiredSamples} & UShort\_t                   & \textit{Number of expired samples in the GAGG.}      \\
\textit{VETO\_ValidEntries}   & UChar\_t                    & \textit{Number of valid entries in the VETO.}        \\
\textit{GAGG\_BackBufferUsed} & UChar\_t                    & \textit{Number of samples in the GAGG's backbuffer.} \\
\textit{CAL\_Mode}            & UChar\_t                    & \textit{Flag for whether CAL mode was used.}         \\
\bottomrule

\end{tabular}%
}
\caption{Types of information contained within the TTree called EventData}\label{table:ROOT}
\end{table}

\begin{table}[H]
\centering
\resizebox{\textwidth}{!}{%
\begin{tabular}{@{}ccc@{}}
\toprule
\textit{\textbf{Histogram}}   & \textit{\textbf{Purpose}}                                                   \\ \midrule
\textit{DELAY\_PULSES}        & \textit{ADC values in GAGG where a delay pulse was found.}                  \\
\textit{PROMPT\_PULSES}       & \textit{ADC values in GAGG where a corresponding prompt pulse was found.}   \\
\textit{GAGG\_DELAYS}         & \textit{The corresponding time difference between prompt and delay pulses.} \\
\textit{VETO\_TIMES}          & \textit{Time values where a VETO pulse was found.}                          \\
\textit{GAGG\_ADC}            & \textit{Full histogram of all GAGG ADC values.}                             \\
\textit{VETO\_ADC}            & \textit{Full histogram of all Veto ADC values.}                             \\
\bottomrule

\end{tabular}%
}
\caption{Histograms provided along with the raw telemetry information.}\label{table:Hists}
\end{table}

\begin{multicols}{2}

\end{multicols}


\section{NuSOL CubeSat Script Generator}\label{sec:scriptgen}

To assist with mission control, a program was created to generate the scripts automatically with an intuitive Graphical User Interface (GUI) rather than manually creating the commands by hand. The GUI utilizes the GTK3 \cite{ref:gtk3} toolkit created in part by the Glade \cite{ref:glade} application, and is designed to run on both Windows and Linux systems, though Windows users will need an extra toolkit called MSYS2 \cite{ref:msys2}. 

\begin{figure}[H]
\begin{center}
\resizebox{\textwidth}{!}{
    \includegraphics[]{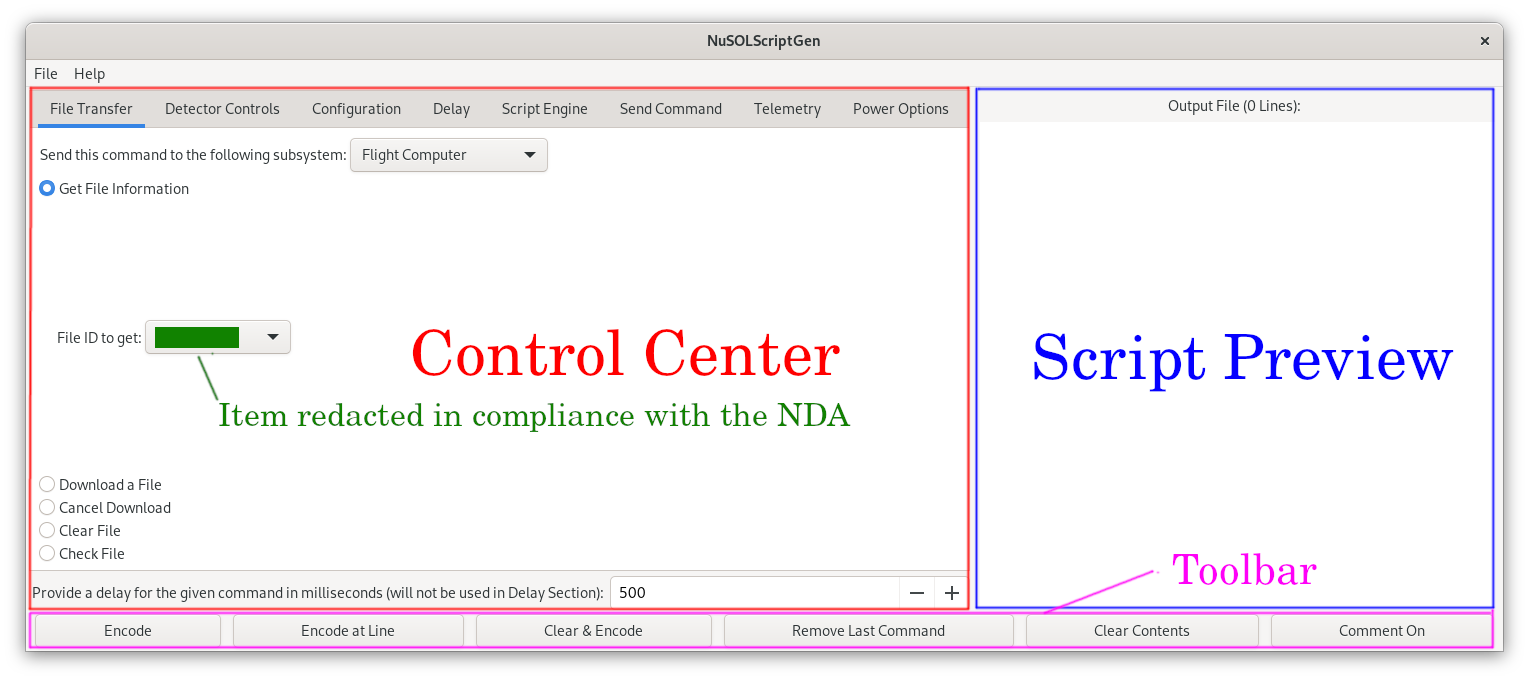}
}
\end{center}
\caption{Initial window of the NuSOL Script Generator; highlighted by layout. Some information has been redacted in green due to a Non-Disclosure Agreement (NDA) being in place for the data directly within the CubeSat itself. }\label{fig:scriptgenpreview}
\end{figure}

This script generator is split into three important sections as shown in Figure \ref{fig:scriptgenpreview}: Control Center for choosing the necessary settings and input in red, Script Preview to view the current script contents in blue, and the Toolbar for generating or modifying the script with a magenta outline (none of these outlines show up in the software). Commands can be selected by first choosing one of the tabs related to the type of command one would want to send, and then choosing the actual command to set the parameters.

Finally, there is a menu bar at the top to provide the options for saving the script file, specially under the \texttt{File} option. It is then up to the personnel to save the modifications to the current script if a file was already selected, or to make a new script file with the given contents. 

\section{Planned Daemon Features}\label{sec:future}
While the daemon is currently implemented, there are two features that are planned to be implemented by the time of launch. The first planned feature in the process of being rolled out is an automated backup daemon along with a a restore daemon that will create redundant copies of the database on both another on-campus machine and a machine off-campus. 

The second planned feature to be rolled out is an automated email system that will send messages over email to the relevant personnel about the status of the daemon. This system will send out status emails at the daily and weekly level, and will be capable of handling various errors on the processing and the database level through priority alerts outside of business hours. 


\section*{Acknowledgments}

    We wish to thank NASA for providing the necessary funding for not just the development of both pieces of software, but also for the SNAPPY CubeSat Mission through NASA in Kansas's Grant No. 80NSSC20M0109 as well as the NIAC Program's Grant No. 80NSSC21K1900. In addition, we would like to thank NanoAvionics and Marshall Space Flight Center for providing us the documentation (provided under NDA) to help decode the data and turn it into useful information for neutrino analysis.
    
\newpage


\printbibliography
\end{document}